# Irradiation of Nanostrained Monolayer WSe$_2$ for Site-Controlled Single-Photon Emission up to 150 K

Kamyar Parto, Kaustav Banerjee, and Galan Moody*

Department of Electrical and Computer Engineering, University of California, Santa Barbara, CA 93106

*E-mail: moody@ucsb.edu

**Abstract**

**Quantum-dot-like WSe$_2$ single-photon emitters have become a promising platform for future on-chip scalable quantum light sources with unique advantages over existing technologies, notably the potential for site-specific engineering. However, the required cryogenic temperatures for the functionality of these sources have been an inhibitor of their full potential. Existing strain engineering methods face fundamental challenges in extending the working temperature while maintaining the emitter's fabrication yield and purity. In this work, we demonstrate a novel method of designing site-specific single-photon emitters in atomically thin WSe$_2$ with near-unity yield utilizing independent and simultaneous strain engineering via nanoscale stressors and defect engineering via electron-beam irradiation. Many of these emitters exhibit exciton-biexciton cascaded emission, purities above 95%, and working temperatures extending up to 150 K, which is the highest observed in van der Waals semiconductor single-photon emitters without Purcell enhancement. This methodology, coupled with possible plasmonic or optical micro-cavity integration, potentially furthers the realization of future scalable, room-temperature, and high-quality van der Waals quantum light sources.**

## Introduction

As the second quantum era emerges in the 21$^{st}$ century, what were once considered as tangential quantum effects are instead harnessed in new quantum technology. Quantum light or single-photon emitters (SPEs) are the heart of many quantum photonic applications[1] such as quantum communication, quantum cryptography, quantum metrology, and linear optical quantum computing. Current state-of-the-art semiconducting single-photon sources[2] – such as InAs self-assembled quantum dots – encounter challenges in scalability and integration with mature photonic technologies due to difficulties associated with their deterministic positioning and growth. In recent years, a new source[3–7] of quantum light has emerged in two-dimensional (2D) materials, including hexagonal boron nitride and transition metal dichalcogenides (TMDs) such as MoSe$_2$ and WSe$_2$. Monolayer WSe$_2$, in comparison to conventional III-V semiconducting quantum dots, offers several advantages, including a high photon extraction efficiency due to its atomic thickness, the potential for scalable, site-controlled and deterministic manufacturability[8–10], and ease of integration with mature photonic technologies via simple transfer methods[11], which has made them an intriguing contender for future technologically relevant quantum light sources.

However, a longstanding challenge facing 2D WSe$_2$ SPEs lies in their low working temperature and thermal instability. The ultra-sharp emission lines associated with SPEs that appear ~50-200 meV below the free exciton of WSe$_2$ are observable only at cryogenic temperatures and quench above 30 K due to low confinement potential and quantum yield[4,12]. Recent studies have attributed the origins of these SPEs to localized intervalley defect excitons that form when the excitonic energies are lowered and hybridize with a valley symmetry-breaking defect state at highly strained regions[13,14] (**Fig.1**). This reveals the pivotal role of both strain and defects in the creation of SPEs. Therefore, engineering high-quality SPEs with high working temperatures ideally requires placement of only a single valley symmetry-breaking defect to minimize the non-radiative recombination and achieve high quantum yield and residing at the center of a strain field to achieve high confinement potential. Earlier methods to deterministically engineer SPEs in 2D TMDs have relied solely on strain engineering methods such as nanopillars or nano-indentation, which in tandem with the high intrinsic defect density of the 2D materials used, would guarantee the presence of at least one defect site in the strained regions. While such methods can achieve a decent fabrication yield, the high intrinsic defect density of the 2D flakes will limit their quantum yield and thermal stability. Therefore, to simultaneously ensure high yield and high thermal stability, it is essential to decouple the strain and defect engineering processes, which has not yet been demonstrated.

In this work, by using electron-beam (e-beam) irradiation as a controllable method to induce structural defects in WSe$_2$, along with engineering strain fields using dielectric nanopillar structures (**Fig.1**), site-specific SPEs in WSe$_2$ are engineered with high yield, high single-photon purity, and possible working temperatures up to 215 K. This sets the precedent for the highest working temperature of SPEs observed in 2D TMDs without Purcell enhancement (See SI.1 benchmark table).

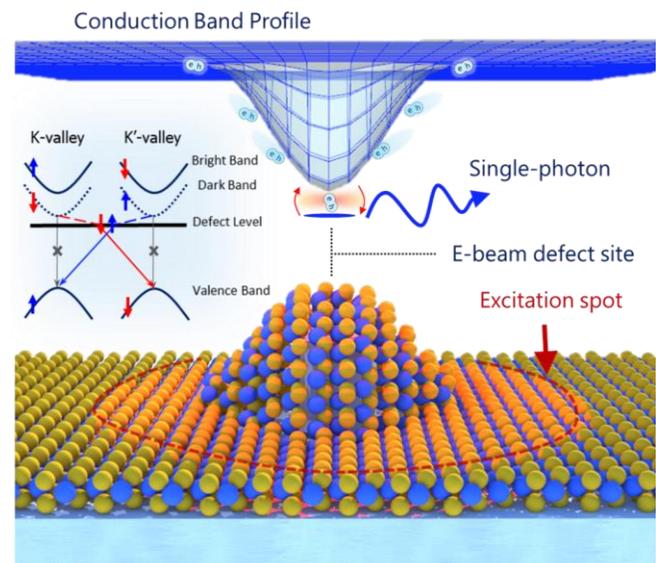

**Fig. 1: Illustration of a strain and defect engineered WSe$_2$ single-photon emitter.** The top denotes the conduction band profile across the flake. The bottom portrays a flake of WSe$_2$ strained over a dielectric nanopillar. The inset figure shows the detailed band-diagram at top of the nanopillar. Delocalized excitons are generated through an excitation laser with a significantly larger spot size compared to the nanopillar size. Note that the efficient radiative decay of these excitons is optically prohibited due to the spin-valley rules of WSe$_2$ (dark excitons, see inset figure). Excitons sense the strain field and funnel into the low potential regions at the top of the nanopillar. The overlap between the confined excitonic states (not shown in the figure) and an e-beam engineered defect near the nano-pillar allows for the formation of localized intervalley defect excitons. Intervalley defect excitons break the valley selectivity and efficiently recombine, giving rise to the bright single-photon emission line as seen in the inset figure.

## Results

A monolayer WSe$_2$ sample encapsulated by h-BN was prepared using dry transfer techniques (see Methods). Subsequently, to engineer the necessary strain profile across the 2D flake, the stack was transferred onto a Si/SiO$_2$ substrate array with a predefined nanopillar dielectric array with height and diameter of 200 nm and 150 nm, respectively, which was chosen as an optimal aspect ratio for deterministically introducing a single emitter per site without piercing through the materials[8]. Photoluminescence spectra of the WSe$_2$ residing on the nanopillars were taken at T = 5 K. Interestingly, no sharp emission peaks that could be associated with single-photon emitters were observed. In contrast, only slightly redshifted neutral exciton (X) and charged exciton (X$^-$) peaks were detected. This is consistent with previous studies[9] and we attribute the small spectral shift to the fact that the laser spot size is much larger compared to the strained region. Overall, this provides evidence that in high-quality encapsulated samples in the absence of any appreciable density of intrinsic defects at the strained sites, single-photon emitters cannot form solely due to the strain in our materials. Next, to deterministically create crystalline defects at the desired positions, the top of the nanopillars were irradiated with an electron beam with tightly focused spot size (<10 nm) and 100 keV accelerating voltage (**Fig. 2a**). E-beam irradiation has been successfully used to induce structural defects[15,16] and phase transformation[17] in 2D TMDs. It is also a well-studied defect engineering method for h-BN SPEs[18]. Other irradiation methods, such as helium ion beam irradiation has also recently demonstrated the ability to create SPEs in 2D MoS$_2$[19].

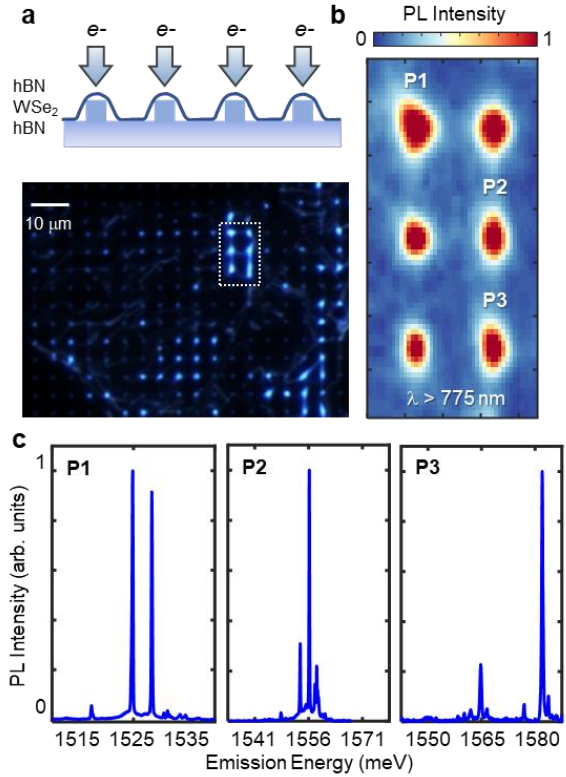

**Fig. 2**: **Site-controlled quantum emitter array via strain and defect engineering.** **(a)** Schematic of the fabricated structure. The bottom figure shows the dark-field image of the sample. Bright emission is observed from the top of the pillars due to enhanced exciton funneling. **(b)** Integrated Photoluminescence map for wavelengths above 775 nm, showing the bright emission associated with single-photon emitters. **(c)** Photoluminescence spectrum captured from different pillar sites showing the sharp emission lines associated with SPEs. Note that many of the sharp emission lines appear in pairs which indicates the possible formation of exciton-biexcitons.

**Fig. 2a** (bottom panel) presents the dark-field optical image of the final assembled structure where the brightly illuminated regions correspond to the nanopillar sites. Integrated photoluminescence (PL) maps **(Fig. 2b)** of six irradiated nanopillars (white dashed box region in **Fig. 2a**) were taken at 5 K for energies below 1.6 eV and provides evidence for the formation of bright emission lines below the WSe$_2$ free exciton. **Fig. 2c**, shows the PL spectrum at the location of 3 different irradiated nanopillars. The neutral and charged excitonic peaks have now been replaced by ultra-sharp emission lines associated with SPEs in these materials, which appear at the energy interval of 1515-1580 meV. The overall picture can be explained well by a recently proposed model[13,14]: Initially, without the presence of valley symmetry-breaking defects, only the neutral and charged excitons spectra are observed. By inducing defects within the strain profile using e-beam irradiation, free excitons funnel to the new intervalley defect sites and recombine with high-rates that give rise to the bright SPE emission lines as described in **Fig.1**.

With a closer look at **Fig. 2c**, it is readily observable that sharp emission lines appear to form in pairs with an energy spacing of about 3-5 meV. The features of these pairs resemble those of an exciton-biexciton where the inherent fine-structure splitting results in two doublets, but the sign of the splitting is reversed in each pair as a consequence of the emission cascade (**Fig. 3a**). The polarization-resolved PL spectrum of the pairs further corroborates the radiative cascade (**Fig. 3b**). The integrated intensities of the pairs show distinct sub-linear and super-linear characteristics as previously observed for exciton-biexciton emissions in WSe$_2$[20] (**Fig. 3c**). Similarly, the time-resolved PL (SI.2) also shows that the decay time of the exciton feature is almost 1.5 times higher than the biexciton, which is comparable with the previously observed dynamics of biexciton-exciton pairs in WSe$_2$[20]. Note that the physical origin of these exciton-biexciton-like features is still ambiguous; previous studies[13] have shown that such features can also be attributed to the hybridization of defect states with localized excitons. Finally, second-order correlation measurements were performed using a Hanbury Brown and Twiss setup[21] that confirms both emission lines act as single-photon sources with purities as high as 90% (**Fig. 3d**). It is worth mentioning that this is the first demonstration of site-specific creation of biexciton features in 2D TMDs.

Engineering high purity and reproducible single-photon emitters with high yield are a requirement for any scalable photonic technology. By decoupling the strain and defect engineering in our nanopillar irradiation process, we were able to achieve a success rate of over 85% in engineering single-photon emitters per site per run. Note that, since e-beam irradiation is a repeatable process, subsequent irradiation and dosage optimization processes can be leveraged to achieve near-unity-yield at each site. **Fig. 4.** represents the statistical evidence of the single-photon emitters fabricated using our method. The spectral purity of many of the emitters was measured to be above 95% (with an average purity of 92%, **Fig.4a**) with average linewidths of 75 µeV and zero-field splitting of 760 µeV. In comparison with previous demonstrations, our method achieves the highest average purity for SPEs in TMDs and it can be expected to achieve even higher purities using resonance excitation.

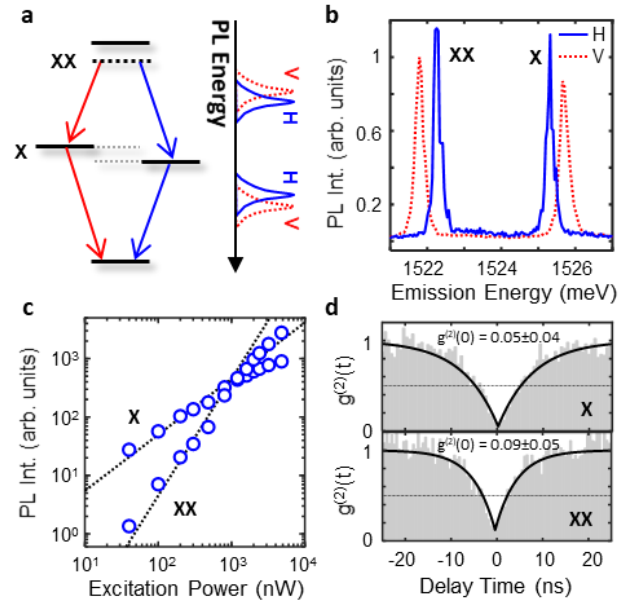

**Fig. 3: Single-emitter photoluminescence and photon antibunching.** **(a)** Cascaded emisison of biexcition. **(b)** PL Spectrum of single-emitter. Two pairs of coupled doublets are observed. **(c) PL intensity vs Excitation Power:** The localized defect exciton shows distinct behavior: Exciton (X) intensitiy increases sublinearly and starts to saturate at high powers as expected from a two-level system, whereas the biexciton line (XX) intensity increases superlinearly. **(d)** Second-order correlation measurement. Both X and XX exhibit $g^{(2)}(0) < 0.1$, showing high purity single-photon emission.

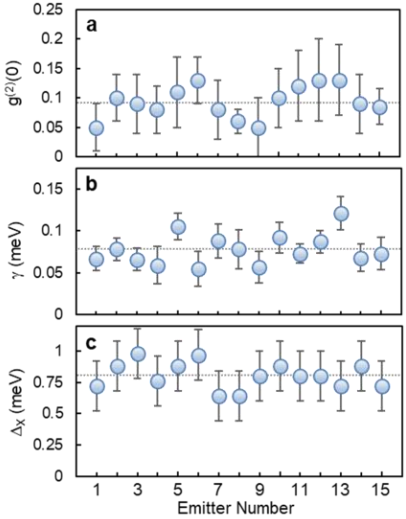

**Fig. 4. Statistics of single-emitter properties.** (**a**) $g^{(2)}(0)$ distribution. Average $g^{(2)}(0)$ is 0.08 which demonstrates the high purities achieved by this method. (**b**) Homogenous linewidth distribution with 75 µeV average. (**c**) Zero-field doublet splitting. Showing the reproducability of the method in engineering SPE's with similar charactristics.

As discussed previously, one of the main disadvantages of WSe$_2$ single photon emitters is that their PL intensity and single-photon purity quench at temperatures above 30 K. One of the main contributors to this effect is the non-radiative scattering of excitons before reaching the single-photon emitter site[12]. In our process, non-radiative recombination is minimized by using high quality exfoliated 2D WSe$_2$ with low defect density (evident from the lack of the SPEs before irradiation of the nanopillars) along with h-BN encapsulation which acts as a shielding layer that minimizes the surface and roughness scattering. Conversely, the defects were induced in the sample via a controllable process targeting only the desired strained regions. This approach minimizes the non-radiative exciton scatterings due to other impurities and ensures that the exciton only radiatively recombines after funneling into the single-photon emitter defect position, which maximizes the quantum yield of the induced emitters. In addition, the average energy of the SPE emission in this work is approximately 60-100 meV lower than previous reports which suggest the emission is originating from deeper states with higher thermal activation energies within the bandgap. It is worth mentioning that this effect can originate either from more accurate positioning of the defect site within the strain field, which improves the confinement potential, or it can also indicate that the morphology of the e-beam-induced defect states responsible for photon emission is inherently different with respect to the defects responsible for SPE in previous studies.

**Fig. 5a** shows the evolution of the single-photon PL emission from 5 K to 150 K (SI.3 is dedicated to the quantitative analysis of single emitter properties). It is readily apparent that the emission line is distinguishable up to 150 K. The relative redshift of the emission line with an increase of temperature is in line with the reduction in the bandgap following Varshni's empirical relationship (**Fig. 5b**). The evolution of the homogenous linewidth broadening with temperature can be seen in **Fig. 5c**. The Arrhenius plot of the integrated intensity of the SPE versus temperature shows that the integrated intensity quenches to $e^{-1}$ at approximately 215 K (**Fig. 5d**). **Fig. 5e-5f**, shows the second-order correlation measurement of the SPE and demonstrates that the emitters exhibit anti-bunching up to 150 K with $g^{(2)}(0) = 0.27\pm0.05$. Given the high average purity (~75%) of the array of emitters at 150 K (see SI.4) in combination with the Arrhenius data, it can be expected that the single-photon nature of the SPEs can persist up to 215 K. To our knowledge, no raw WSe$_2$ SPEs have been able to operate as a single-photon emitter up to 150K without Purcell enhancement. It can be predicted that our SPE engineering method, in tandem with designs leveraging Purcell enhancement, could potentially achieve functionality at room temp

## Discussion

**Fig.6** quantitatively recaps one plausible mechanism behind the single-photon emission as described earlier in **Figure.1**. The dielectric nanopillar creates a strain profile akin to **Fig.6a** in the 2D flake where the strain is at its maximum around the edges of the nanopillar. The applied strain results in the reduction of the bandgap (**Fig. 6b**), which then creates a potential landscape analogous to **Fig.6a**. This potential leads to the localization of the free exciton states. From e-beam irradiation, defect states appear within the bandgap of the semiconductor in the strained regions. Provided the energy of these defect states at the strained regions is sufficiently close to the conduction band of the localized exciton states, the defects and excitons hybridize and results in the bright single-photon emission peak. While the strain and engineering method outlined in this work demonstrate the interplay between defects and strain, as predicted by the above-mentioned picture, the physical morphology of the defects responsible for SPE remains elusive. There is a lack of consensus of the defects responsible for single-photon emission, but they have been previously attributed to structures including selenium vacancies[13], tungsten centered vacancies[22], oxygen interstitials[23], and anti-site defects[24]. It is also plausible that multiple types of defect may be responsible for the single-photon emission, as long as they break the valley symmetry, and upon application of strain, introduce defect states with favorable energies close to the conduction band.

However, given that the defects in this work were created by e-beam irradiation for which the morphology of resulting defects has been extensively studied, the list of possible suspects can be reduced through a systematic, ab-initio approach. Studies have shown that e-beam irradiation processes primarily contribute to the generation of chalcogen and double chalcogen vacancies due to a lower knock-off energy[15].

**Fig. 5. Statistics of single-photon emission up to 150 K.** (**a**) Evolution of the PL spectrum as function of the temperature. (**b**) Red shift of the SPE line with temperature increase due to the reduction of the bandgap following Varshni's empirical relationship. (**c**) Homogenous linewidth broadening due to the increase of temperature. (**d**) Arhenious plot of the integrated intensity. The PL peak starts to quench around 150 K and reaches $e^{-1}$ at 215 K. (**e**) Second-order correlation measurement $g^{(2)}$ at T=5 K and (**f**) at 150 K. exhibiting $g^2(0) < 0.3$, evidencing that the single-photon nature persists up to our maximum measurement temperature of 150 K and likely persists beyond 200 K.

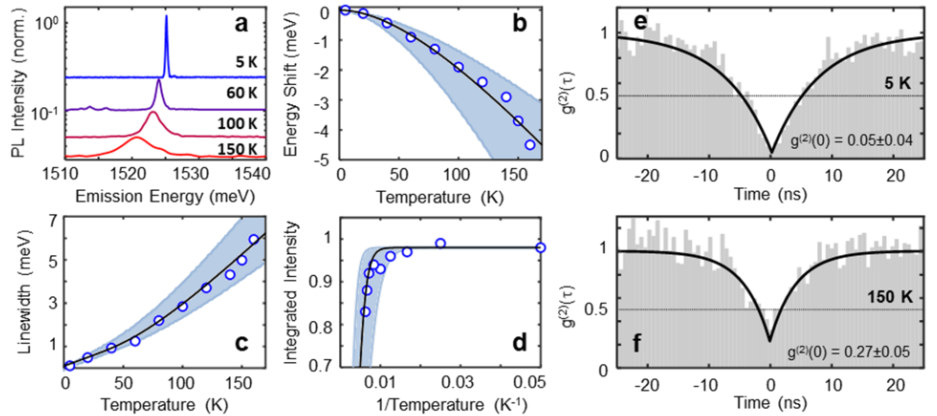

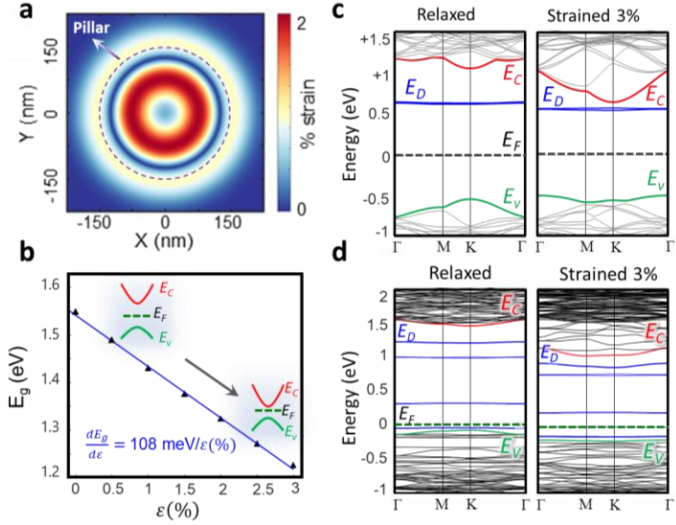

**Fig. 6. DFT Calculations of strained defect bandstructure (a)** Simulated strain profile at the nanopillar location. **(b)** Bandgap as a function of applied biaxial tensile strain. Note that the conduction band movement is larger with respect to the valence band. **(c)** Evolution of the band diagram of selenium vacancy with strain. $E_D$ denotes the localized defect level. Upon the application of strain, $E_c$ and $E_D$ come into proximity. **(d)** Evolution of the band diagram of $WSe_6$ pore complex versus strain. After the application of strain, the energy spacing between the conduction band and its closest defect-level decreases. Note that the Fermi-level energy position depends on the size of the supercell.

Furthermore, prolonged exposure of a focused e-beam spot can also lead to the destabilization of the transition metal bond, which causes the transition metal to migrate away from the site, creating more complex vacancy sites such as pore vacancies[25,26]. In the case of $WSe_2$, the creation of rotational trefoil defect complexes under e-beam irradiation has also been observed[27]. Therefore, we chose to study five probable defect complexes that might form during the e-beam process, namely selenium single vacancy, selenium double vacancy, tungsten vacancy, pore vacancy, and trefoil vacancies (See SI.5 for full details).

DFT calculations were performed using Synopsis QuantumATK[28] package by incorporating Perdew-Burke-Ernzerh (PBE) variant of Generalized Gradient Approximation[29] (PBE-GGA) exchange-correlation functional (see Methods for details). Note that given the common bandgap underestimation problem in DFT, along with the uncertainties in the exciton binding energies and the magnitude of the confinement potential, a quantitative solution to this problem seems impractical. However, here our main criterion is the relative movement of the conduction band and defect levels with respect to each other upon application of strain, which can be assessed properly using ab-initio simulation.

**Fig.6c** shows the dynamics of selenium single vacancy levels with applied biaxial tensile strain. It is readily observable that the defect levels appear at favorable energies close to the conduction band and move to the proximity of the conduction band with strain. However, both selenium vacancies and selenium double vacancies are prone to passivation by oxidation[30]. Which given the fact that the SPEs in this work, similar to previous studies, remain functional after many days and temperature cycles, makes these defects an unlikely candidate. Conversely, tungsten vacancies do not introduce energies in the vicinity of $E_C$ and, therefore, are not energetically favorable (SI.5). $WSe_6$ vacancy complex – pore vacancies – on the other hand, form nanopores similar to those observed in $MoS_2$[25] and $WS_2$[26] after geometry relaxation (SI.5). These vacancies also introduce energies close to $E_C$, which move towards the conduction band upon further application of strain (**Fig.5d**). Thus, it is plausible that more complex vacancy structures created by e-beam irradiation, such as nanopores, are responsible for the SPEs observed in the e-beam irradiated sites.

## Conclusion

In conclusion, by decoupling the strain and defect engineering process in the design of $WSe_2$ single-photon emitters, we were able to achieve near-unity yield in deterministic positioning of quantum emitters with high purity that can preserve their single-photon nature to well above 150 K. This method was also successful at deterministic engineering of exciton-biexciton-like features with a radiative cascade that can pave the way for the realization of entangled photon-pair sources. It can be expected that by utilizing the methodology described in this work, and integration with plasmonic or microphotonic cavities, our work sets the stage for future scalable elevated-temperature high-quality 2D $WSe_2$ quantum light sources.

## Methods

**Sample Preparation:** The nanopillar array was fabricated with high-resolution electron beam lithography following the procedures described in Ref.[9]. Briefly, a silicon wafer with thermal oxide was cleaved into a 1 $cm^2$ chip followed by a solvent clean. The chip was spin-coated with hydrogen silsesquioxane (HSQ) resist diluted with methyl isobutyl ketone (MIBK). After a 5-minute bake at 90 °C, pillar arrays and chip alignment markers were defined via electron beam lithography and then developed in a 25% tetramethyl ammonium hydroxide (TMAH) solution and rinsed with methanol. A final rapid thermal anneal step at 1000 °C in oxygen converts the HSQ pattern into $SiO_2$. The nanopillars have a nominal diameter of 150 nm and height of 200 nm.

We used a dry viscoelastic transfer technique to create the hBN-encapsulated $WSe_2$ structure. $WSe_2$ flakes (2D Semiconductors) were exfoliated onto an unpatterned preparation $SiO_2$/silicon substrate for identification and subsequent transfer. Next a thin flake of hBN is exfoliated onto the transfer stamp, which was used to exfoliate and transfer $WSe_2$ monolayers from thicker flakes on the preparation substrate. Finally, the $WSe_2$/hBN stack was transferred to an hBN layer exfoliated on the patterned nanopillar array, which ensures that the $WSe_2$ layer is encapsulated by the hBN and free of any contaminants or residues during and after the transfer process. Quantum emitters were created using a 100 keV, < 10 nm spot size electron-beam to expose the sample at the location of each nanopillar with ~$10^6$ electrons/$\mu m^2$.

**Optical Spectroscopy:** For steady-state micro-photoluminescence (PL) measurements, the sample temperature was held fixed between 5 K and 150 K. The emitters were excited non-resonantly with a continuous-wave 633 nm laser focused to a ~1 μm full-width at half-maximum spot size using a long working distance 0.7 numerical aperture microscope objective. The backscattered PL was collected and measured with a spectrometer and TE-cooled charge-coupled device with a spectral resolution of ~30 μeV. For time-resolved PL measurements, a 40 MHz, 635 nm pulsed laser diode was used as the excitation source. The PL was spectrally filtered with ~1 nm bandwidth and detected using a superconducting nanowire single-photon detector and time-tagging electronics with ~40 ps detector temporal resolution. The PL measurements were performed at various average excitation powers ranging from 10 nW to 10 μW. Polarization-resolved spectra were acquired with a half-wave plate and linear polarizer before the spectrometer.

**DFT Calculations:** HGH pseudopotentials with Tier4 basis sets, 3×3×1 and 3×3×1 Brillouin zone k-point sampling, 200 Rydberg density mesh cut-offs and a 0.02 eV/Å maximum force constant were used for geometry optimizations and DFT calculations. GGA-HGH Tier4 setup has been widely used in literature to probe the physics of 2D TMDs. $WSe_2$ supercell was geometrically relaxed using the setup discussed above to 0.02 eV/ Å maximum force constant. Next, depending on the size of the defect complex, a 5×5 or 7×7 supercell was constructed, followed by the removal of the appropriate atoms. Next, the geometry was optimized again by constraining the supercell along the in-plane direction to achieve the target strain and allowing for the defective supercell to relax to its final configuration.

## Acknowledgment

We gratefully acknowledge support from the UC Santa Barbara NSF Quantum Foundry funded via the Q-AMASE-i program under award DMR-1906325. G.M. acknowledges support from the National

Science Foundation (ECCS-2032272). K.P. was supported by the ARO (grant W911NF1810366). Use was made of computational facilities purchased with funds from the National Science Foundation (CNS-1725797) and administered by the Center for Scientific Computing (CSC). The CSC is supported by the California NanoSystems Institute and the Materials Research Science and Engineering Center (MRSEC; NSF DMR 1720256) at UC Santa Barbara.

**Author Contributions**

G.M. conceived the experiments. All authors contributed towards all aspects of this research.

**Competing interests**

The authors declare no competing interests.